\UseRawInputEncoding 
\documentclass[preprint,superscriptaddress,amsmath,longbibliography,amssymb,prl]{revtex4-2}

\usepackage{graphicx}
\usepackage{dcolumn}
\usepackage{bm}
\usepackage{multirow}
\usepackage{array}
\usepackage[usenames,dvipsnames]{xcolor}
\definecolor{nblue}{rgb}{0.0, 0.0, 1.0}
\definecolor{magenta}{rgb}{0.79, 0.08, 0.48}
\usepackage[colorlinks,linkcolor=nblue,urlcolor=magenta,citecolor=magenta,plainpages=false,pdfpagelabels,breaklinks]{hyperref}

\begin{document}
\title{Phase Transitions and Superconductivity in Ternary Hydride Li$_2$SiH$_6$ at High Pressures}

\author{Sixuan Wu}
\affiliation{School of Science, Nanjing University of Posts and Telecommunications, Nanjing 210023, China}
\author{Bin Li}
\email{libin@njupt.edu.cn}
\affiliation{New Energy Technology Engineering Laboratory of Jiangsu Province and School of Science, Nanjing University of Posts and Telecommunications, Nanjing 210023, China}
\author{Zhi Chen}
\affiliation{State Grid Zhenjiang Power Supply Company, Zhenjiang 212000, China}
\author{Yu Hou}
\affiliation{College of Electronic and Optical Engineering, Nanjing University of Posts and Telecommunications, Nanjing 210023, China}
\author{Yan Bai}
\affiliation{College of Electronic and Optical Engineering, Nanjing University of Posts and Telecommunications, Nanjing 210023, China}
\author{Xiaofeng Hao}
\affiliation{College of Electronic and Optical Engineering, Nanjing University of Posts and Telecommunications, Nanjing 210023, China}
\author{Yeqian Yang}
\affiliation{College of Electronic and Optical Engineering, Nanjing University of Posts and Telecommunications, Nanjing 210023, China}
\author{Shengli Liu}
\affiliation{New Energy Technology Engineering Laboratory of Jiangsu Province and School of Science, Nanjing University of Posts and Telecommunications, Nanjing 210023, China}
\author{Jie Cheng}
\affiliation{New Energy Technology Engineering Laboratory of Jiangsu Province and School of Science, Nanjing University of Posts and Telecommunications, Nanjing 210023, China}
\author{Zhixiang Shi}
\affiliation{School of Physics, Southeast University, Nanjing 211189, China}
\date{\today}

\begin{abstract}

  We predicted a new ternary hydride Li$_2$SiH$_6$ at high pressures. A systematic structure search in Li$_2$SiH$_6$ compound reveals novel stable phases with intriguing electronic and phonon properties. It is found that Li$_2$SiH$_6$ is dynamically stable from ambient pressure up to 400 GPa with three novel phases: P312, P$\bar{3}$, and P$\bar{6}$2m. The calculation of electron-phonon coupling combined with Bardeen-Cooper-Schrieffer's argument indicates that this compound may be a candidate for high $T_c$ superconductors under high pressures. In particular, the maximum $T_c$ of $P\bar{6}2m$-Li$_2$SiH$_6$ at 400 GPa reaches 56 K. These findings may pave the way for obtaining room temperature superconductors in dense hydrogen-rich compounds.
\end{abstract}

\maketitle

\section{Introduction}
The search for high-temperature superconducting materials has always been a hot topic in the field of condensed matter physics. According to the traditional Bardeen-Cooper-Schrieffer (BCS) theory\cite{bardeen1957theory}, the superconducting transition temperature is directly proportional to the Debye temperature, and the Debye temperature is inversely proportional to the mass, so lighter elements may have higher superconducting transition temperatures, such as hydrogen. However, solid hydrogen is an insulating molecular crystal under ambient pressure, and there are strong covalent bonds in hydrogen molecules, so it is hard to obtain superconducting hydrogen under ambient pressure. In order to achieve superconductivity, conditions such as external pressure are needed. As a basic thermodynamic variable, pressure can change the distance between atoms of matter, so that atoms can be rearranged, and the crystal and electronic structure can be modulated. High pressure can very effectively shorten the distance between atoms, increase the overlap of adjacent electron orbits, and then change the interaction and electronic structure between atoms/molecules, forming high-pressure new phases with new structures and properties that are difficult to form under conventional conditions. Generally, in a sufficiently high pressure environment, the band gap will be narrowed, and the energy bands will overlap, which can transform the non-metallic state into the metallic state. Wigner and Huntington proposed that insulating hydrogen molecules can be transformed into metallic hydrogen under high pressure, showing a metallic state\cite{wigner1935possibility}. Ashcroft proposed chemical preloading\cite{ashcroft2004hydrogen}, that is, non-hydrogen elements in hydrogen-rich compounds have an interaction effect on hydrogen elements in the lattice. Hydrogen-rich compounds can be metallized at a lower pressure than pure hydrogen, and such hydrogen-rich compounds may be potential room temperature superconductors. Therefore, hydrogen-dominated metal alloys may be potential high-temperature superconducting materials under high pressure, because of the high Debye temperature. In recent years, hydrogen sulfide (H$_3$S), discovered under high pressure, set a high-temperature superconductivity record due to its superconducting transition temperature of 203 K \cite{Duan2014,Drozdov2015}, and subsequent theoretical studies and experimental demonstrations found that lanthanum hydride compounds are at a high pressure of 170 GPa, with a superconducting transition temperature of 250 K \cite{Somayazulu2019,Drozdov2019}. This is the highest superconducting transition temperature confirmed by experiments on binary hydrides so far, which has further promoted the research on the superconductivity of hydrogen-rich compounds.

In the "hydride rush" of searching high-temperature superconductors, the theoretical research of hydrides has stepped into the forefront. First-principles calculations based on density functional theory have given a series of superconductivity predictions about hydrogen-rich compounds, and proposed a new family of superconducting hydrides with a unique cage structure\cite{Wang6463,Peng2017,Liu2018}. The metal atoms such as calcium or lanthanum are located in the center of the hydrogen cage. In these high $T_c$ superconducting hydrides, doping elements metalize hydrogen at relatively lower pressures, through "chemical pre-compression". Based on the characteristics of "chemical pre-compression", some ternary hydrogen storage materials are considered as candidates for high-temperature superconductors. Ternary metal hydrides are a convenient and valuable system for studying the metallization and superconductivity of metal hydrides because they can be synthesized under mild conditions and recovered under ambient pressure\cite{Meng2019}. Some ternary hydrides have been found in theoretical and experimental work, e. g. Fe$_2$SH$_3$ (T$_c$ = 0.3 K, 173 GPa)\cite{Zhang2016}, YS$_4$H$_4$ (T$_c$ = 20 K, 200 GPa)\cite{Grishakov2019} and YSH$_5$ (T$_c$ = 0.7 K, 300 GPa)\cite{Chen2019}, BaReH$_9$ (T$_c$ = 7 K above 100 GPa)\cite{Muramatsu2015}. A recent experimental work also reported a room-temperature superconductivity with T$_c$  of 288 K in the C-S-H system at 267 GPa\cite{Snider2020}.

Silicon and carbon belong to the same group IV elements. It is controversial that silicon atoms will bond with four H atoms to form SiH$_4$ molecular units in the Mg-Si-H system similar to MgCH$_4$. However, since the nucleus of silicon is larger than that of carbon, the strength of the Si-H bond is weaker than that of the C-H bond. This may promote the dissociation of the Si-H bond, resulting in a completely different crystal structure, and there will be new changes under high pressure. A typical example is the MH$_4$ (M = Si\cite{Eremets2008}, Ge\cite{Gao2008}, and Sn\cite{Tse2007}) system, where the difference in bonding results in greatly different structures and properties. In addition, recent studies have found that lithium atom doping helps stabilize the structure of the Au-H system \cite{Rahm2017}, because under high pressure, the Au and H systems are unstable, but the doping of lithium has a stabilizing effect, which may originate from the low electron negative of lithium. Based on first-principles calculations, by adding lithium to the poor superconducting MgH$_{12}$, it is found that the superconducting transition temperature of Li$_2$MgH$_{16}$ reaches 472 K under 250GPa \cite{Sun2019}, indicating that it can be doped with lithium atoms, carrying ultra-high superconducting transition temperature. Zhang et al. performed structure searching simulations on the Li-Si-H system at pressure range of 50-350 GPa\cite{Zhang2020}. LiSi$_2$H$_9$ and LiSiH$_8$ are predicted to become good phonon-mediated superconductors with estimated $T_c$ of 54 and 77 K at 172 and 250 GPa, respectively. Li$_2$SiH$_6$ compound is predicted with relatively low symmetry (P-1 and Cmcm) and insulating. While another study on Li$_x$XH$_6$ predicted LiPH$_6$ a potential high-temperature superconductor with a $T_c$ of 150-167 K at 200 GPa\cite{shao2019ternary}. Boeri et.al also revealed a metallic Li$_2$BH$_6$ phase which is thermodynamically stable down to 100 GPa with a critical temperature of $\sim$ 100 K\cite{Kokail2017}. Thus, the stable phases and superconductivity of Li$_x$XH$_6$ needed further exploration. In this paper, we focus on Li$_2$SiH$_6$ compound, searching for structures and superconductivity under high pressures. We found a series of phase transitions and a candidate high $T_c$ superconductors $P\bar{6}2m$-Li$_2$SiH$_6$ with the maximum $T_c$ of 56 K. Our results will further explore the superconductivity in the Li-Si-H system and may help to guide future experiments.

\section{Methods}
 To determine the high-pressure structures, we used our machine-learning-based crystal prediction code CRYSTREE~\cite{crystree}. CRYSTREE can learn the structural optimization procedure and output optimized structure using multi regression methods including random forest regressor, ridge regressor and ExtraTrees regressor, et al.~\cite{scikit-learn}. The results of structures are checked by evolutionary crystal structure prediction method USPEX~\cite{uspex}. The electronic structure calculations with high accuracy for the stable structures were performed using the full-potential linearized augmented plane wave (FP$-$LAPW) method implemented in the WIEN2K code.~\cite{Wien2k}  The Wu-Cohen generalized gradient approximation (WC-GGA)~\cite{GGA} was applied to the exchange-correlation potential calculation. The plane-wave cutoff was defined by $RK_{max}=4.0$, where $R$ is the minimum LAPW sphere radius and $K_{max}$ is the plane-wave vector cutoff. The phonon calculations were carried out by using a density functional perturbation theory (DFPT) approach through the Quantum-ESPRESSO program~\cite{QE}. Ultrasoft pseudopotentials are used in the calculation with the cutoffs chosen as 60 Ry for the wave functions and 600 Ry for the charge density. The electronic self-consistent calculation was performed over a $16\times16\times16$ $k$-point mesh. Dynamical matrices and the electron-phonon couplings were calculated on a $4\times4\times4$ $q$-point grid. A dense $24\times24\times24$ grid was used for evaluating an accurate electron-phonon interaction matrix.

\section{Results}

\begin{figure}[htbp]
	\begin{center}
		\includegraphics[scale=0.5]{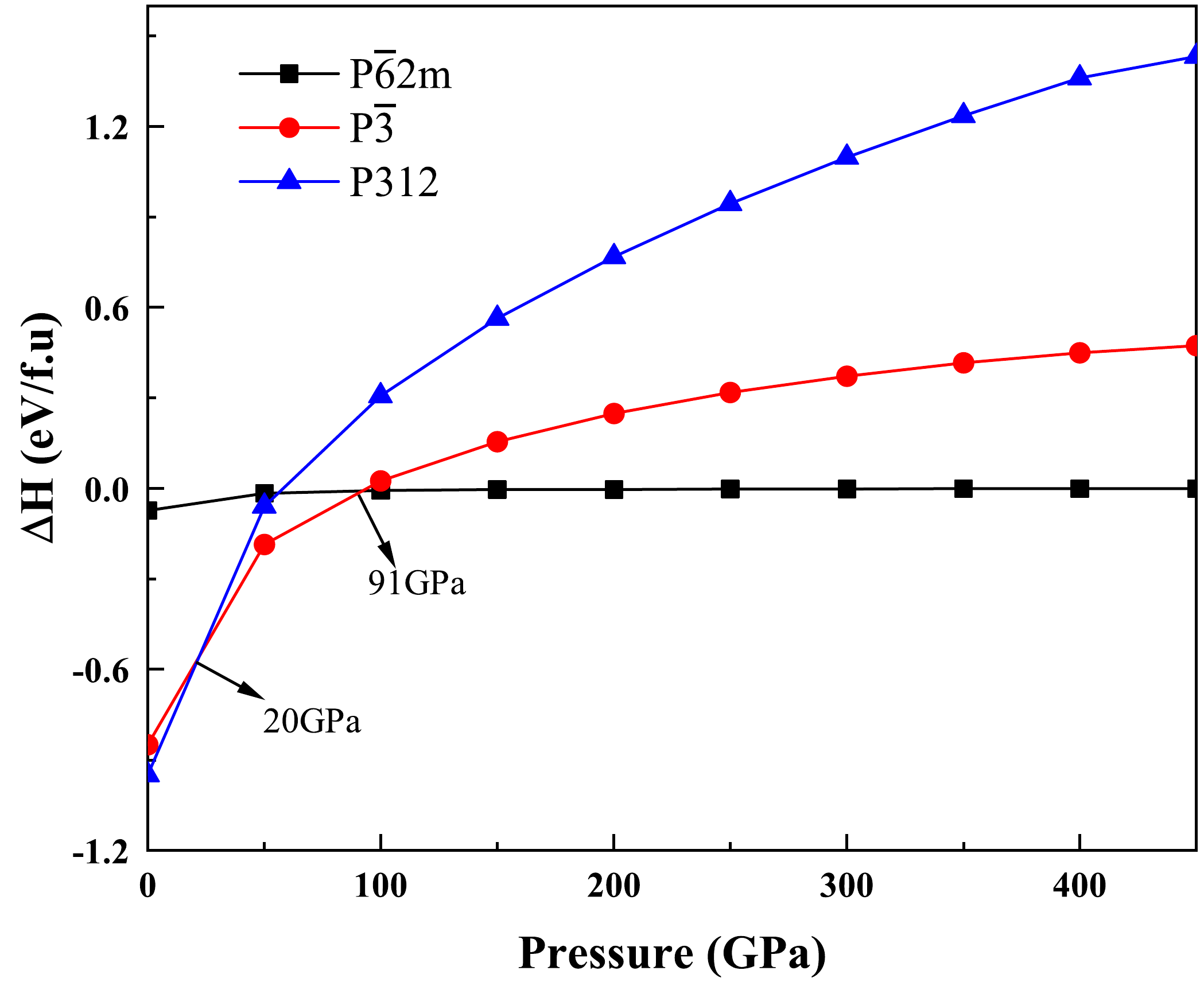}\\
		\caption{The enthalpies per formula unit of P312 and P$\bar{3}$ phases as a function of pressure, referenced to P$\bar{6}$2m phase of Li$_2$SiH$_6$. }
		\label{enthalpy}
	\end{center}
\end{figure}

The crystal structures of Li$_2$SiH$_6$ are predicted at 0, 100, 200, 300 and 400 GPa, respectively, with system sizes containing up to 4 formula units (f.u.) per simulation cell. Three energy-competitive phases P312, P$\bar{3}$ and P$\bar{6}$2m were found through our structure prediction. As shown in figure \ref{enthalpy}, from ambient pressure up to 20 GPa, the more stable phase is P312 phase (space group NO.149), then P$\bar{3}$ phase (space group NO.147) becomes a more stable phase until 91 GPa, following a phase transition to P$\bar{6}$2m phase (space group NO.189) up to the highest pressure we studied of 400 GPa. Li$_2$SiH$_6$ undergoes a series of phase transitions from trigonal symmetry (P312 and P$\bar{3}$ phases) to hexagonal symmetry (P$\bar{6}$2m phase).

\begin{figure}[htbp]
	\includegraphics[scale=0.5]{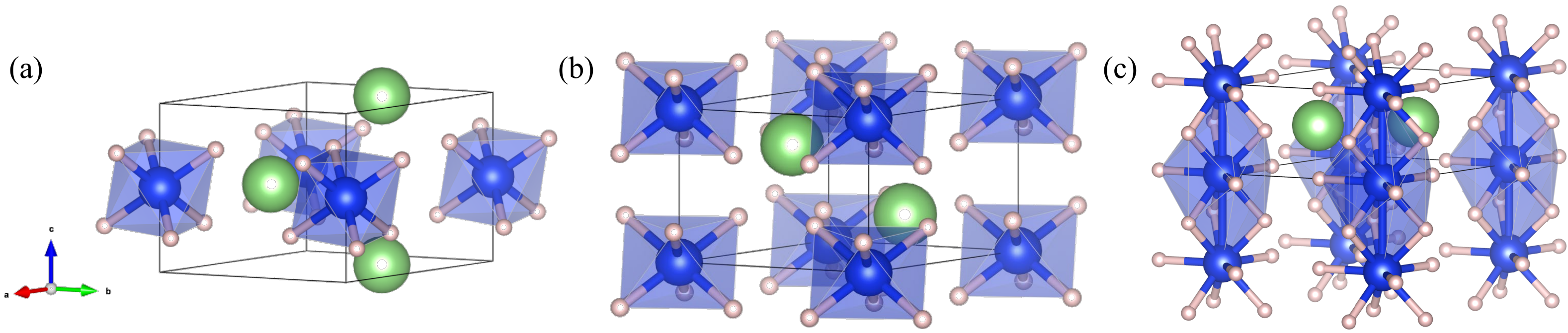}\\
	\caption{Predicted crystal structure of Li$_2$SiH$_6$ phases. Lithium atoms are represented by large green balls, silicon atoms are represented by blue balls, and hydrogen atoms are represented by small pink balls. The crystal structure of the predicted phase is generated using VESTA software.~\cite{vesta}} \label{structures}
\end{figure}

All of the crystal structures of the stable Li$_2$SiH$_6$ compounds at 0 GPa, 50 GPa, and 300 GPa are shown in Fig. \ref{structures}. The low-pressure phases P312 and P$\bar{3}$ contain Si-H6 units in the cell, with lithium atoms inserted between these units. For high-pressure P$\bar{6}$2m phase, there are Si-H9 units with discrete lithium atoms. With the increasing pressure, more hydrogen atoms trend to condensate around silicon atom. This is a prerequisite for a higher superconducting transition temperature in hydrides. In P312 and P$\bar{3}$ phases, Si-H6 units construct regular octahedrons, in which Si-H bond in P312 phase is 1.599 \AA\ and the length of Si-H bonds of P$\bar{3}$ phase is 1.472 \AA. When the pressure further increases, the SiH$_n$ units change from the original octahedrons to 14-sided polyhedrons, and more hydrogen atoms condensate and bonded with central Si atom. Each Si atom in P$\bar{6}$2m structure is connected to nine H atoms with Si-H bonds ranging from 1.417 \AA\ to 1.480 \AA. A Si-Si bond with a bond length of 2.014 \AA\ is also formed in this phase at 300 GPa.

\begin{figure}[htbp]
		\includegraphics[scale=0.4]{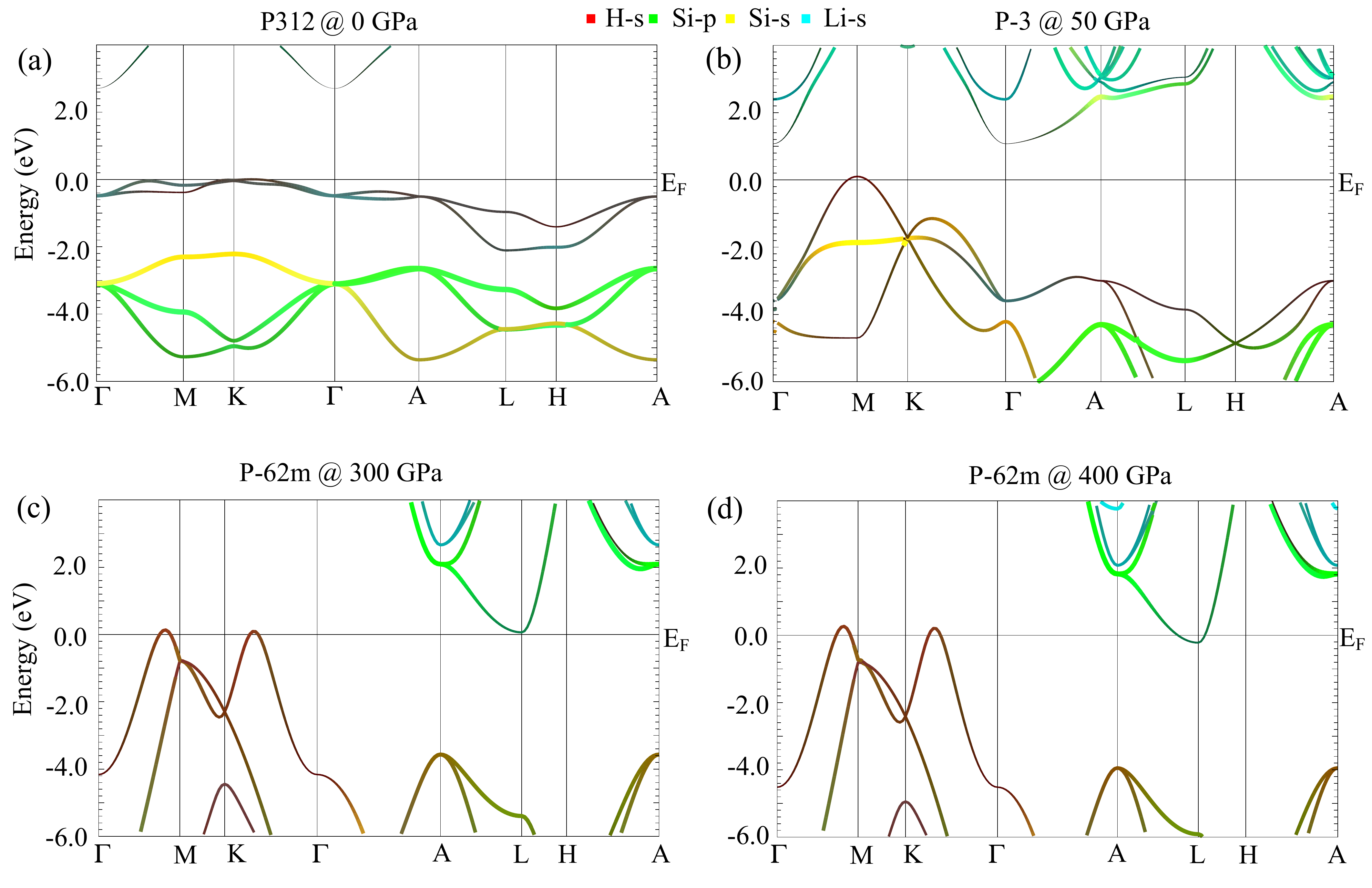}\\
		\caption{Electronic band structure for each phase. (a) P312 phase at 0 GPa, (b) P$\bar{3}$ phase at 50 GPa, (c) P$\bar{6}$2m at 300 GPa. (d) P$\bar{6}$2m at 400 GPa. The Fermi level is set to zero. The red, green, yellow, and cyan color represent the contributions of H-s, Si-p, Si-s and Li-s orbitals to the density of states.} \label{bandstructures}
\end{figure}

\begin{figure}[ht]
	\includegraphics[scale=0.5]{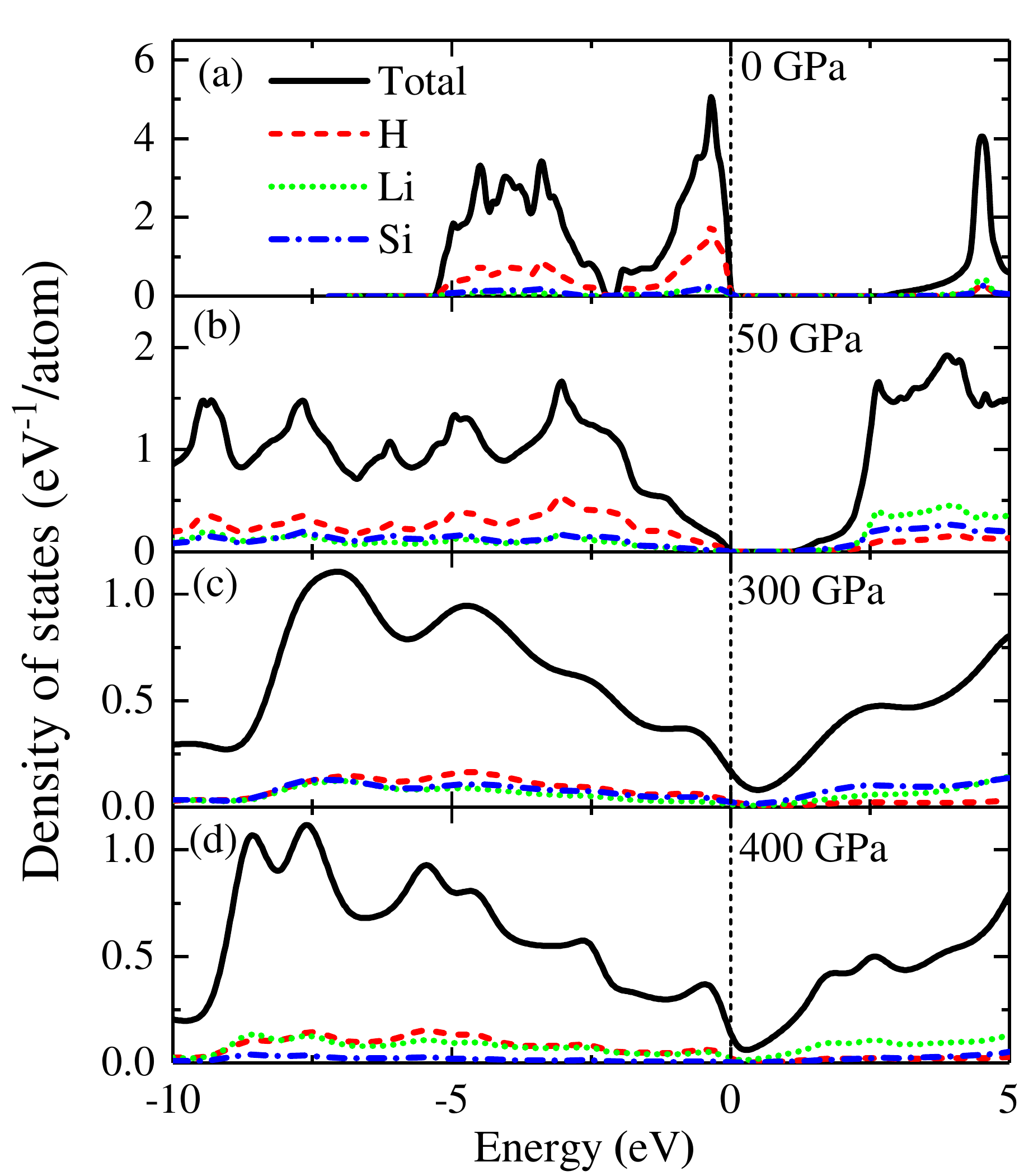}\\
	\caption{The density of states of P$\bar{6}$2m, P$\bar{3}$ and P312 phases at different pressures.} \label{dos}
\end{figure}

Figure \ref{bandstructures} shows the calculated electronic band structure of Li$_2$SiH$_6$ for P312, P$\bar{3}$ and P$\bar{6}$2m phases. Band structure for P312 phase at 0 GPa are shown in Fig. \ref{bandstructures}(a), there is no energy band that crosses the Fermi level ($E_F$), which demonstrates P312 phase an insulator at ambient pressure. As shown in Fig. \ref{bandstructures}(b), with the increasing pressure, the most stable phase changes from P312 to P$\bar{3}$. A valence band slightly crossing $E_F$ at 50 GPa which shows a weak metal characteristic. P$\bar{6}$2m becomes the most stable phase at 91 GPa. As shown in figure \ref{bandstructures} A hole-type band cross through the Fermi level between $\Gamma-M$ and $K-\Gamma$ at 300 GPa. When pressure further increase to 400 GPa, there are one electron-type band and two hole-type bands crossing the Fermi energy. This may help enhance the electron-phonon interaction. The total and projected density of states (DOS) are shown in figure \ref{dos}. In low-pressure phases P312 at ambient pressure and P$\bar{3}$ at 50 GPa, the band gap are around 3 eV and 1.2 eV, respectively. While in high-pressure phase P$\bar{6}$2m, the DOS at the Fermi level $E_F$ show remarkable mentality  with 0.17 eV$^{-1}$/atom at 300 GPa and 0.15 eV$^{-1}$/atom at 400 GPa. The contribution to the DOS of each element shows that the orbitals are highly hybridized.

\begin{figure}[ht]
	\includegraphics[scale=0.5]{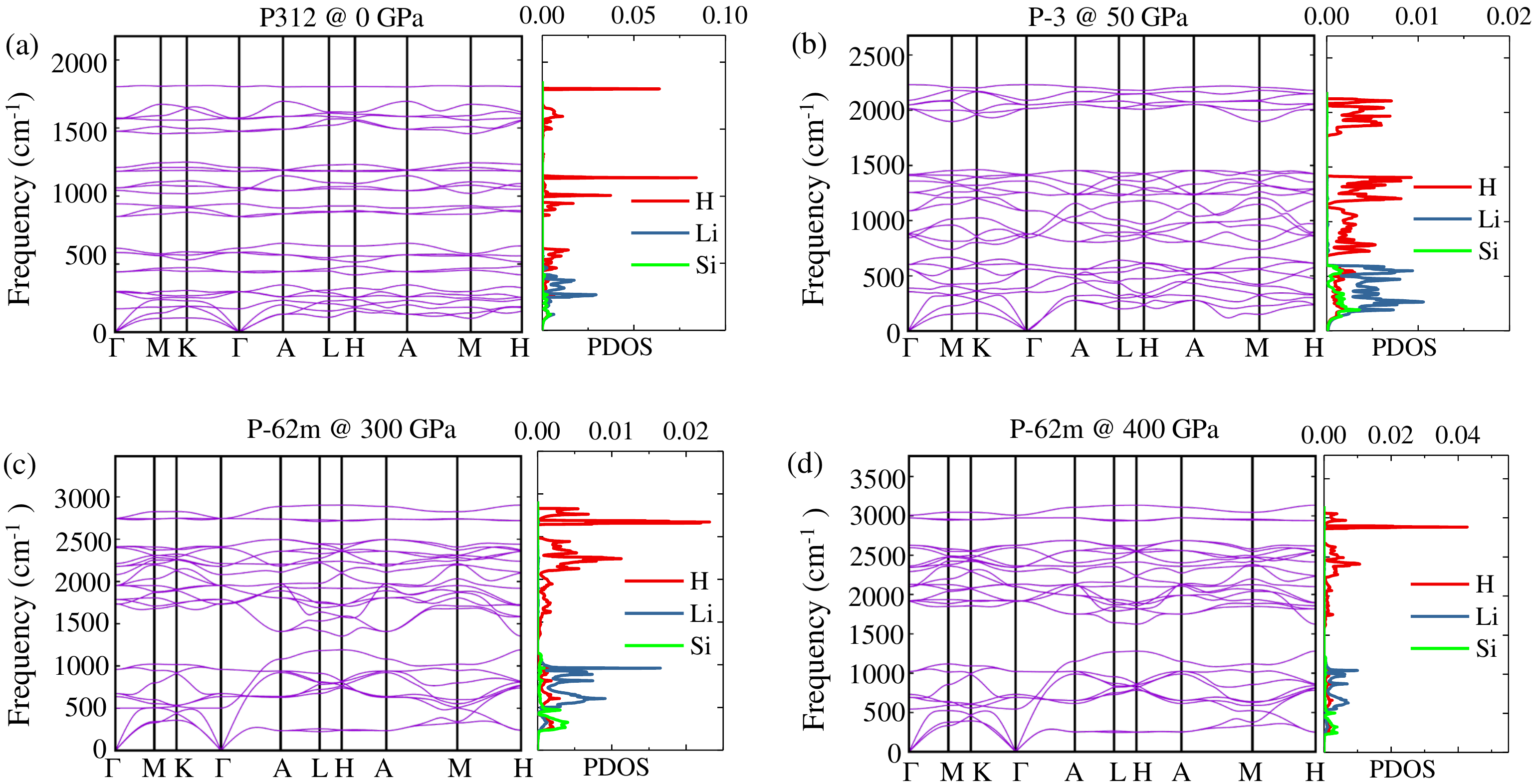}\\
	\caption{The phonon dispersion curves and partial phonon density of states of P312, P$\bar{3}$ and P$\bar{6}$2m phases, under different pressures.} \label{phonon}
\end{figure}

Figure \ref{phonon} show the phonon spectra and partial phonon density of states (PDOS) of Li$_2$SiH$_6$ under (a) P312 at 0 GPa, (b) P$\bar{3}$ at 50 GPa, (c) P$\bar{6}$2m at 300 GPa and (d) P$\bar{6}$2m at 400 GPa, respectively. No soft phonon mode can be observed in all phases at the pressures we studied, indicating that these structures are dynamically stable. We also preformed symmetry analysis on the phonon modes. For P312 phase, its point group is $D_3$. The irreducible representation can be written as $\Gamma=2A'_1 \bigoplus 2A'_2 \bigoplus A''_1 \bigoplus 4A''_2 \bigoplus 7E' \bigoplus 2E''$, with $3A''_2 \bigoplus 6E'$ infrared active modes and $2A'_1 \bigoplus 6E' \bigoplus 2E''$ Raman active modes. For P$\bar{3}$ phase, its point group is $C_{3i}$, The irreducible representation can be written as $\Gamma=4A_g \bigoplus 5A_u \bigoplus 4^1E_g \bigoplus 5^1E_u \bigoplus 4^2E_g \bigoplus 5^2E_u$, with $4A_u \bigoplus 4^1E_u \bigoplus 4^2E_u$ infrared active modes and $4A_g \bigoplus 4^1E_g \bigoplus 4^2E_g$ Raman active modes. For P$\bar{6}$2m phase, its point group is $D_{3h}$. The irreducible representation can be written as $\Gamma=2A'_1 \bigoplus 2A'_2 \bigoplus A''_1 \bigoplus 4A''_2 \bigoplus 7E' \bigoplus 2E''$, with $3A''_2 \bigoplus 6E'$ infrared active modes and $2A'_1 \bigoplus 6E' \bigoplus 2E''$ Raman active modes. The PDOS show that, the vibrations of hydrogen atoms dominate the high-frequency region, while vibrations of lithium and silicon mainly distribute in the medium and low-frequency regions, respectively. It is consistent with the ordering of elements' masses, the heavier atoms prefer vibrations with lower frequencies. We also noted that, with the increasing pressure, the maximum phonon frequency changes from $\sim$ 1700 cm$^{-1}$ at 0 GPa, to $\sim$ 2300 cm$^{-1}$ at 50 GPa, then to $\sim$ 2800 cm$^{-1}$ at 300 GPa and $\sim$ 3000 cm$^{-1}$ at 400 GPa. It will help to enhance the Debye temperature and logarithmic average frequency $\omega$$_{log}$, which is directly proportional to the superconducting critical temperature.

\begin{table}[ht]
	\renewcommand{\arraystretch}{1.2}
	\setlength{\tabcolsep}{8mm}
	\caption{The logarithmical averaged phonon frequency $\omega_{log}$, electron-phonon coupling ($\lambda$), and superconducting $T_{c}$ in Li$_2$SiH$_6$ at different pressures.}
	\begin{tabular}{l*{4}c}
		\hline
		Structure           & P(GPa) & $\omega$$_{log}$ (K) & $\lambda$ & $T_c$ (K) ($\mu$$^*$=0.10) \\\hline
		P312-Li$_2$SiH$_6$  & 0      & 784.45              & 0.003     & 0.00                  \\
		P$\bar{3}$-Li$_2$SiH$_6$   & 50     & 880.68              & 0.23      & 0.01                  \\
		P$\bar{6}$2m-Li$_2$SiH$_6$ & 100    & 1692.10             & 0.04      & 0.00                  \\
		                    & 200    & 2059.12             & 0.07      & 0.00                  \\
		                    & 300    & 1134.88             & 0.73      & 43.60                 \\
		                    & 400    & 1143.95             & 0.82      & 56.33                 \\

		\hline
	\end{tabular}\label{Tc}
\end{table}

TABLE I collects predicted superconducting properties of Li$_2$SiH$_6$ phases. The calculated logarithmic average phonon frequency $\omega_{log}$, electron-phonon coupling parameters $\lambda$, and superconducting critical temperatures $T_c$ at various pressures are listed. $T_c$ in the range of 100 and 400 GPa rise upon compression. We use the Allen and Dynes modified McMillan equation~\cite{Allen1975,McMillan1968} to estimate the superconducting temperature $T_c$ at different pressures.
\begin{equation}
\omega_{log} = exp(\frac{2}{\lambda}\int_{\omega_{min}}^{\omega_{max}}\frac{d\omega}{\omega}\alpha^2F(\omega)ln(\omega))
\end{equation}
The Coulomb pseudopotential $\mu^*$ is assumed to be 0.1 for all materials.  Along with the calculated logarithmic average frequency $\omega$$_{log}$, the resultant $T_c$ of P$\bar{6}$2m-Li$_2$SiH$_6$ reaches maximum value of 56 K at 400 GPa, which is the maximum pressure we studied.

\section{Conclusions}
We searched stable structures of the Li$_2$SiH$_6$ compound in the pressure range from 0 to 400 GPa using the machine-learning-based crystal structure prediction algorithm, and predicted new stable phases of P312, P$\bar{3}$ and P$\bar{6}$2m at different pressures. The maximum $T_c$ value of Li$_2$SiH$_6$ phases obtained by first-principles calculations are 56 K in high pressure P$\bar{6}$2m phase. This study complements the investigation of a series of high-$T_c$ superconducting ternary hydrides.

\section{Acknowledgments}
This work is sponsored by the National Key R\&D Program of China (Grant No. 2018YFA0704300), the National Natural Science Foundation of China (Grants No. U1932217, 12175107), open program from the National Lab of Solid State Microstructures of Nanjing University (Grant No. M32025), and NUPTSF (Grant No. NY219087, NY220038).

\vspace{10pt}
\noindent\textbf{Author contributions:}
{S.X.W wrote the paper with input from all the co-authors}.

\vspace{10pt}
\noindent\textbf{Competing financial interests:}
The authors declare no competing financial interests.

%


\end{document}